# Image Encryption Based On Gradient Haar Wavelet and Rational Order Chaotic Maps

Sodeif Ahadpour[1], Yaser Sadra[*1,2], Meisam Sadeghi[2]

[1]Faculty of Sciences, University of Mohaghegh Ardabili, Ardabil, Iran.

[2]Department of Mathematics & Computer sciences, Shandiz Institute of Higher School, Mashhad, Iran.

[*]Corresponding author e-mail: sadra@ijtpc.org

## Abstract

Haar wavelet is one of the best mathematical tools in image cryptography and analysis. Because of the specific structure, this wavelet has the ability which is combined with other mathematical tools such as chaotic maps. The rational order chaotic maps are one of clusters of chaotic maps which their deterministic behaviors have high sensitivity. In this paper, we propose a novel method of gradient Haar wavelet transform for image encryption. This method use linearity properties of the scaling function of the gradient Haar wavelet and deterministic behaviors of rational order chaotic maps in order to generate encrypted images with high security factor. The security of the encrypted images is evaluated by the key space analysis, the correlation coefficient analysis, and differential attack. The method could be used in other fields such as image and signal processing.

## Keywords

## 1. Introduction

Nowadays, signal and image processing are one of important fields in communication sciences. One of methods for the analysis is data discrete transforms [1, 2, 3, 4, 5, and 6]. The data discrete transforms could be used in many fields such as image compression, data compression, encrypted data, and so on [7, 8, 9, 10, and 11]. The compressions are considered in the branches of the multimedia which need the speedy digital video, high performance, and audio capabilities [16, 17, 18, and 19]. Discrete Wavelets such as Haar wavelet are hierarchy of the data discrete transforms. Recently, merit of the discrete Wavelets is much attention in among Scientists and researchers [12, 13, 14, and 15]. Haar wavelet transforms are a method for the compression which is used as wavelet-based compression algorithms in the JPEG-2000 standard [19, 20, and 21]. Because of the simplicity of structure of Haar wavelet transforms, image encryption was discussed less based on their. In previous work, we propose gradient Haar wavelet in order to upgrade Haar wavelet transforms in the image encryption [22]. Here, we first review gradient Haar wavelet [22] and rational order chaotic maps [23, 24]. Then, a novel algorithm is proposed based on their which can be using in image and data encryption.

## 2. Preliminaries

### 2.1 Gradient Haar Wavelet

The simplest possible wavelets are Haar wavelet. They most widely used in the various sciences. Gradient Haar transform or gradient Haar wavelet transform is modified the Haar wavelet transforms which was proposed in 1909 by Alfred Haar [6]. Based on the multiresolution analysis, Haar transforms have the scaling function and the Haar wavelet with various shifts and stretches. The scaling function is as following [5, 6]:

$$\phi(x) = \begin{cases} 1 & 0 \leq x < 1 \\ 0 & other\ wise \end{cases} \quad (1)$$

and the Haar wavelet is as follows:

$$\psi(x) = \begin{cases} 1 & 0 \le x < \tfrac{1}{2} \\ -1 & \tfrac{1}{2} \le x < 1 \\ 0 & otherwise \end{cases} \quad (2)$$

Their graphs are given in Figure 1. In other mathematical words, the scaling function and wavelet function are as following [5, 6]:

$$\phi(x) = \sum_{l \in z} p_l\, \phi(2x - l) \quad (3)$$

or the set of functions

$$\phi_{j,l}(x) = \{2^{\tfrac{j}{2}} \phi(2^j x - l); l \in Z\}$$

where for any $j \in Z$, $l$ determines the position of $\phi_{j,l}(x)$ along the x-axis, $j$ determines $\phi_{j,l}(x)$'s width. Then

$$\psi(x) = \sum_{l \in z}(-1)^l\, \overline{p_{1-l}}\, \phi(2x - l) \quad (4)$$

that the $\widetilde{p_l}$ coefficients in here are called scaling function coefficients and are as following:

$$p_l = 2 \int_{-\infty}^{\infty} \phi(x)\, \overline{\phi(2x-l)}\, dx \quad \text{and} \quad \widetilde{p_l} = 2^{\tfrac{-1}{2}} p_l .$$

The Haar scaling function and wavelet function are defined as ($p_0 = p_1 = 1$) [6]

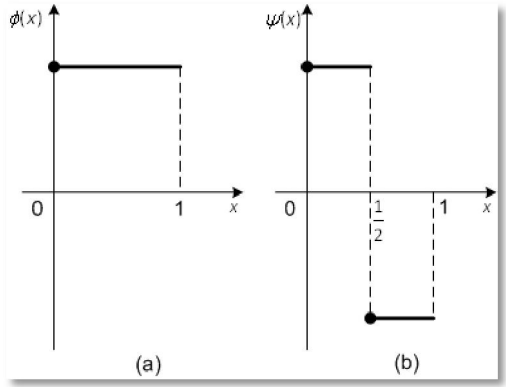

Fig.1. Graph of $\phi(x)$ and $\psi(x)$ of Haar wavelet.

Since the scaling function coefficients of Haar wavelet are constant values [6] ($p_0 = p_1 = 1$), the Haar wavelet transform is used less in cryptography. Therefore, researchers usually use of the Haar wavelet transforms in cryptography based on the hybrid methods [25, 26, and 27]. In previous work [22], we introduced gradient Haar wavelets transform which the scaling function and the wavelet function are not constant values. In other word, the scaling function of gradient Haar wavelet changed as a sloping step function [22]:

$$\phi(x) = \begin{cases} \lambda\left(x - \tfrac{1}{2}\right) + 1 & 0 \le x < 1 \\ 0 & other\ wise \end{cases} \quad (5)$$

Where $\lambda$ is line slop and $-2 \le \lambda \le 2$. In result, the gradient Haar wavelet is as follows []:

$$\psi(x) = \begin{cases} \left(\tfrac{\lambda^2}{24} + \tfrac{\lambda}{4} + 1\right)(2\lambda x - \tfrac{\lambda}{2}+1) & 0 \le x < \tfrac{1}{2} \\ -\left(\tfrac{\lambda^2}{24} - \tfrac{\lambda}{4} + 1\right)(2\lambda x - \tfrac{3\lambda}{2}+1) & \tfrac{1}{2} \le x < 1 \\ 0 & otherwise \end{cases}$$
(6)

In other other mathematical word, the scaling function and wavelet function of the gradient Haar wavelet transform are as follows:

$$\phi(x) = \left(\tfrac{\lambda^2}{24} - \tfrac{\lambda}{4} + 1\right)\phi(2x) + \left(\tfrac{\lambda^2}{24} + \tfrac{\lambda}{4} + 1\right)\phi(2x - 1)$$

$$\psi(x) = \left(\tfrac{\lambda^2}{24} + \tfrac{\lambda}{4} + 1\right)\phi(2x) - \left(\tfrac{\lambda^2}{24} - \tfrac{\lambda}{4} + 1\right)\phi(2x - 1)$$

$$p_0 = \tfrac{\lambda^2}{24} - \tfrac{\lambda}{4} + 1,\ p_1 = \tfrac{\lambda^2}{24} + \tfrac{\lambda}{4} + 1 \quad (7)$$

which the functions are the Haar wavelet functions if be $\lambda = 0$. The graphs of the new scaling function and function of wavelet are given in Figure 2.

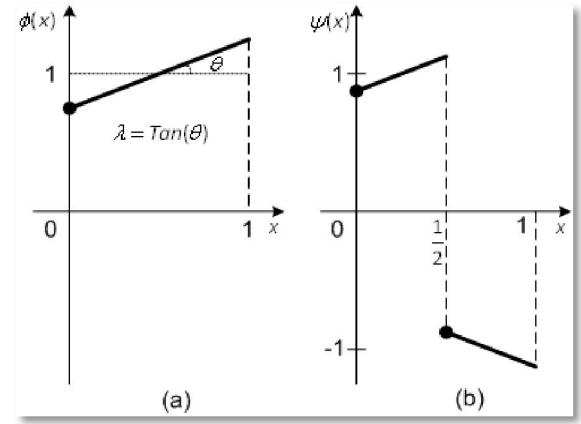

Fig.2. Graph of $\phi(x)$ and $\psi(x)$ of gradient Haar wavelet.

Using the tensor product of one-dimensional wavelets could be created two-dimension gradient Haar wavelet. Consequently, the multiresolution analysis on a two-dimension signal $f_{n,n}$ is as following [6]:

$$f_{n,n} = LL_{n-1,n-1} + LH_{n-1,n-1} + HL_{n-1,n-1} + HH_{n-1,n-1}$$

Where $LL_{n-1,n-1}, LH_{n-1,n-1}, HL_{n-1,n-1},$ and $HH_{n-1,n-1}$ are respectively the linear combinations of $\psi(2^{n-1}x - l)f_{n,n}(x,y)\psi(2^{n-1}y - \hat{l}),\ \psi(2^{n-1}x - l)f_{n,n}(x,y)\phi(2^{n-1}y - \hat{l}),\ \phi(2^{n-1}x -$

$l)f_{n,n}(x,y)\psi(2^{n-1}y-\hat{l})$, and $\phi(2^{n-1}x-l)f_{n,n}(x,y)\phi(2^{n-1}y-\hat{l})$.

Here, L is approximation sequence or low-pass filter in horizontal, and also H is details sequence or high-pass filter in horizontal or vertical directions of the matrix 6, 22]. In the image analysis application, a two-dimension signal $f_{n,n}$ can be a plain image. According to the above, the two-dimension gradient Haar wavelet obtained. An example of a $4 \times 4$ gradient Haar transformation matrix is as following [22]:

$$H_{4\times 4} = \begin{vmatrix} (\widetilde{p_0})^2 & \widetilde{p_0}\widetilde{p_1} & \widetilde{p_1}\widetilde{p_0} & (\widetilde{p_1})^2 \\ \widetilde{p_1}\widetilde{p_0} & (\widetilde{p_1})^2 & -(\widetilde{p_0})^2 & -\widetilde{p_0}\widetilde{p_1} \\ \widetilde{p_1} & -\widetilde{p_0} & 0 & 0 \\ 0 & 0 & \widetilde{p_1} & -\widetilde{p_0} \end{vmatrix} \quad (8)$$

where the $\widetilde{p_l} = \frac{p_l}{\sqrt{2}}$ coefficients are scaling function coefficients. Thus, we obtained a two-dimension gradient Haar wavelet transform with variable scaling function coefficients. The two-dimensional gradient wavelet transform decomposes the plain image to four sub-images (LL, LH, HL, and HH). The LL image can be decomposed to four new sub-images which make a tree of sub-images as shown in Figure 3. The Haar gradient transform can be presented in matrix form $F = GMG^T$ where M is a $n \times n$ plain matrix, G is a $n \times n$ gradient Haar transform matrix, and F is the $n \times n$ resulting transform matrix.

## 2.2 The rational order chaotic maps

The rational order chaotic maps are one of clusters of the one-parameter chaotic maps which their deterministic behaviors have high sensitivity [23]. The rational order chaotic maps are as the ratio of polynomials of degree N [23, 24].

$$\varphi_N(x,a) = \frac{a^2\left(1+(-1)^N F_1^2\left(-N,N,\frac{1}{2},x\right)\right)}{(a^2+1)+(a^2-1)(-1)^N F_1^2\left(-N,N,\frac{1}{2},x\right)}$$

where $N$ is an integer greater than 1. Here, we use the rational order chaotic map with N=2 which is as following:

$$\varphi_2(x,a) = \frac{a^2(2x-1)^2}{4x(1-x)+a^2(2x-1)^2}$$

Hence, the chaotic map can be as a chaotic generator in image encryption:

$$x_{n+1} = \varphi_2(x_n,a) \quad (9)$$

where $\alpha$ is the chaotic control parameter.

## 2.3 Chaotic gradient Haar Wavelet Transform

We introduce chaotic gradient Haar wavelet transform with variable scaling function coefficients which is a new method in image encryption [22]. We focus on the chaotic gradient Haar wavelet in two-dimension and use the rational order chaotic maps to generate $s$. Consequently, we have a chaotic gradient Haar transformation matrix from variable scaling function coefficients($\widetilde{p_l}$). A $4 \times 4$ chaotic gradient Haar transformation matrix is as following

$$H_{4\times 4} = \begin{vmatrix} (\widetilde{p_0(s_1)})^2 & \widetilde{p_0(s_2)}\widetilde{p_1(s_3)} & \widetilde{p_1(s_4)}\widetilde{p_0(s_5)} & (\widetilde{p_1(s_6)})^2 \\ \widetilde{p_1(s_7)}\widetilde{p_0(s_8)} & (\widetilde{p_1(s_9)})^2 & -(\widetilde{p_0(s_{10})})^2 & -\widetilde{p_0(s_{11})}\widetilde{p_1(s_{12})} \\ \widetilde{p_1}(s_{13}) & -\widetilde{p_0}(s_{14}) & 0 & 0 \\ 0 & 0 & \widetilde{p_1}(s_{15}) & -\widetilde{p_0}(s_{16}) \end{vmatrix}$$
(10)

As an example, we obtain a $4 \times 4$ chaotic gradient Haar transformation matrix with proposed control parameters of the rational chaotic maps (Eq.9) as:

$\{x_0 = 0.6, \ \alpha = 2\}$.

Then,

$$\begin{cases} s_1 = 0.143 & s_2 = 0.806 & s_3 = 0.706 & s_4 = 0.451 \\ s_5 = 0.037 & s_6 = 0.960 & s_7 = 0.956 & s_8 = 0.952 \\ s_9 = 0.947 & s_{10} = 0.941 & s_{11} = 0.934 & s_{12} = 0.925 \\ s_{13} = 0.912 & s_{14} = 0.895 & s_{15} = 0.869 & s_{16} = 0.827 \end{cases}.$$

Consequently, we have:

$$H_{4\times 4} = \begin{bmatrix} 0.931 & 0.696 & 0.886 & 0.637 \\ 0.638 & 0.640 & -0.642 & -0.645 \\ 0.806 & -0.809 & 0 & 0 \\ 0 & 0 & 0.814 & -0.821 \end{bmatrix}.$$

We can develop the transformation matrix to a $n \times n$ chaotic gradient Haar transformation matrix.

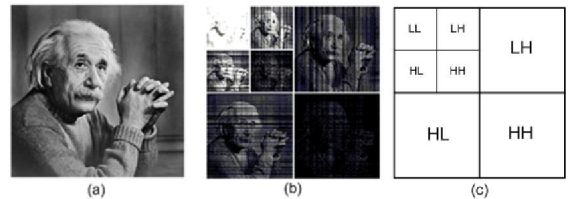

Fig.3. (a) The plain image "Einstein.bmp". (b) Two levels Haar wavelet decomposition of the plain image "Einstein.bmp". (c) Two dimensional 2-leveles gradient Haar wavelet decomposition.

Here, we are made a chaotic gradient Haar wavelet transformation. Using this method, contrast ratio (brightness) of the sub-images has randomly changed in horizontal and vertical (see Fig. 3 (b)).

# 3. Proposed image encryption algorithm

In order to encrypt image, we will describe the proposed encryption method for images based on the chaotic gradient Haar wavelet transforms (CGHWT). Fig. 4 displays the block diagram of the proposed encryption method. In the algorithm just suffice which we use only 1-level 2D wavelet transform.

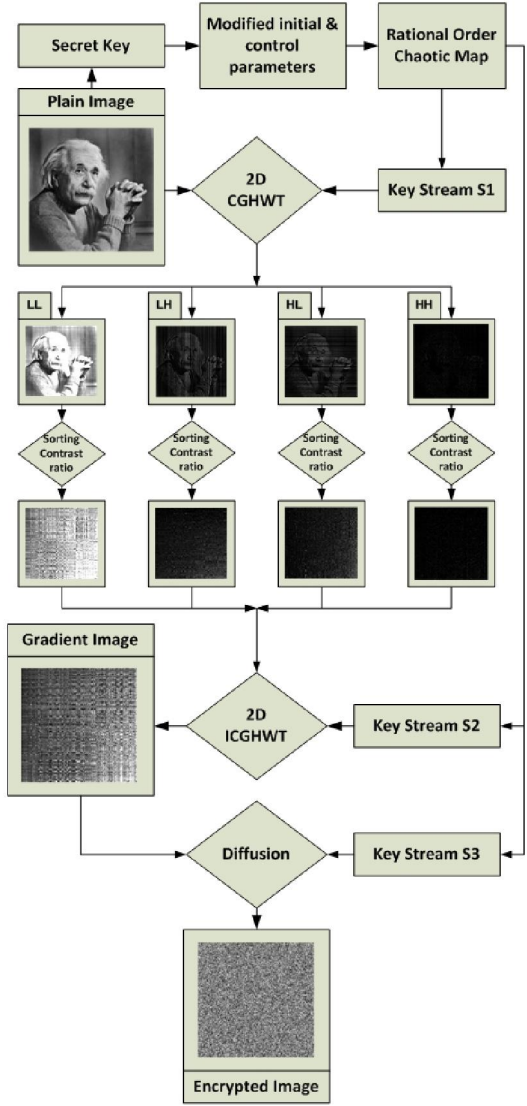

Fig.4. The block diagram of the proposed encryption.

The encrypted image is produced by the following steps:

**Step 1.** Input plain image $P$ with size $m \times n$ and then we produce secret keys based on pixels of the plain image as follows:

$$\lambda_N = mod\left(\sum_{i=1}^{m} P(i,N), \left\lceil \frac{\sum_{i=1}^{m}\sum_{j=1}^{n} P(i,j)}{m \times n} \right\rceil\right) \oplus 255 \quad (11)$$

$$\mu_N = mod\left(\sum_{j=1}^{n} P(N,j), \left\lceil \frac{\sum_{i=1}^{m}\sum_{j=1}^{n} P(i,j)}{m \times n} \right\rceil\right) \oplus 255 \quad (12)$$

Where $\oplus$ represents the bitwise XOR operator and index N is degree of polynomials of the rational order chaotic map. Next, we produce initial and control parameters $x$ and $\alpha$ of the rational order chaotic map based on secret keys $\lambda_N$ and $\mu_N$.

$$x_k = (\lambda_N \oplus \mu_N)/255 \quad (13)$$

$$\alpha_k = N(1 + x_k) \quad (14)$$

Where index $k$ is number of key streams (S**1**, S**2**, and S**3**). Then, the key streams are produced by the initial and control parameters and the rational order chaotic map.

**Step 2.** Produce a 2D chaotic gradient Haar wavelet transform (CGHWT) matrix from variable scaling function coefficients $\{\widetilde{p}_l(S1_t) | t = 1, \dots, m \times n\}$ based on key stream *S1*. Next, decompose plain image with 1-level 2D CGHWT in order to compute the approximation coefficients matrix LL, and details coefficients matrices LH, HL, and HH. The Fig 2 and 3 show which the contrast ratios of the sub-images has randomly changed in horizontal and vertical. The value of average of the contrast ratio is computed to each row and column. Then, the contrast ratios of the sub-images sort once horizontally and one again vertically based on the value of average of them.

**Step 3.** Produce a 2D inverse chaotic gradient Haar wavelet transform (ICGHWT) matrix from variable scaling function coefficients $\{\widetilde{p}_l(S2_t) | t = 1, \dots, m \times n\}$ based on key stream *S2*. Next, apply the 1-level 2D inverse chaotic gradient Haar wavelet transform (ICGHWT) in order to produce gradient image $G$.

**Step 4.** An encrypted image is produced from combination the gradient image and the key stream *S3* $\{S3_t | t = 1, \dots, m \times n\}$ as following:

$$E(i,j) = G(i,j) \oplus [S3_t \times 255]$$

where $\oplus$ represents the bitwise XOR operator. Consequently, an encrypted image $E$ is obtained by this method. It resists every security and statistical attacks due to having many initial and control parameters. The decryption process can be almost the same as the encryption but with reverse steps (see Fig. 5).

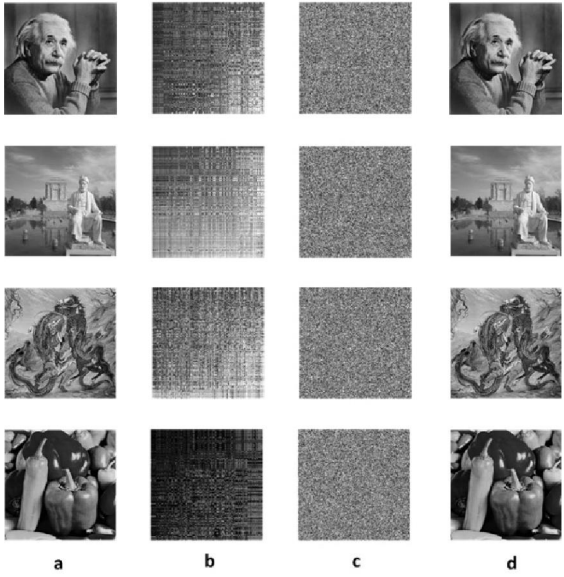

Fig.5. Encryption and decryption of plain images (Einstein, Ferdowsi, Rostam, and Pepper) in the proposed method: (a) plain images; (b) gradient images; (c) encrypted images; (d) decrypted images.

# 4. Experimental results and performance analysis

The most important part of an image encryption is security. Then, a complete analysis should apply to ensure security of the encrypted images of the proposed method. Here, based on performance analysis, we show that the encrypted image are resistant versus various cryptographically attacks.

### 4.1 Key Space

The key space analysis is one of the mathematical methods to check security. The key space should be large enough which encrypted image can resist brute-force attacks. The secret keys should be specific enough to the encryption algorithm [28, 31]. As mentioned in Section 3, the secret keys and key streams are all double precision real numbers. Based on the IEEE floating point standard [35], the computational precision of the 64 bit double precision number is about $10^{-15}$. In this method, there are four initial and control parameters at least in the interval 0 to 1 that were used two times in chaotic gradient Haar wavelet transforms. Hence, the size of secret key is much larger than $2^{398}$. Such a big key space is large enough to defeat brute-force by any super computer today.

### 4.2 Histogram

The histogram is the mathematical method to illustrate pixel's contrast ratio distribution of the images. Here, we use the images with different pixel's contrast ratio distributions to investigate ability of encrypting of the proposed algorithm. The histograms are shown in Figs. 6. The histogram of the encrypted images is substantially different from the histogram of the plain images. As we know, the uniform pixel's contrast ratio distribution is one of the properties of the encrypted image which the encrypted images of the proposed algorithm sufficiently have its. In addition, the table 1 shows that the average pixel intensities for all of the encrypted images are nearly equal and equal to the average value of pixels. Consequently, they do not provide any useful information to perform any statistical analysis attack on the encrypted images.

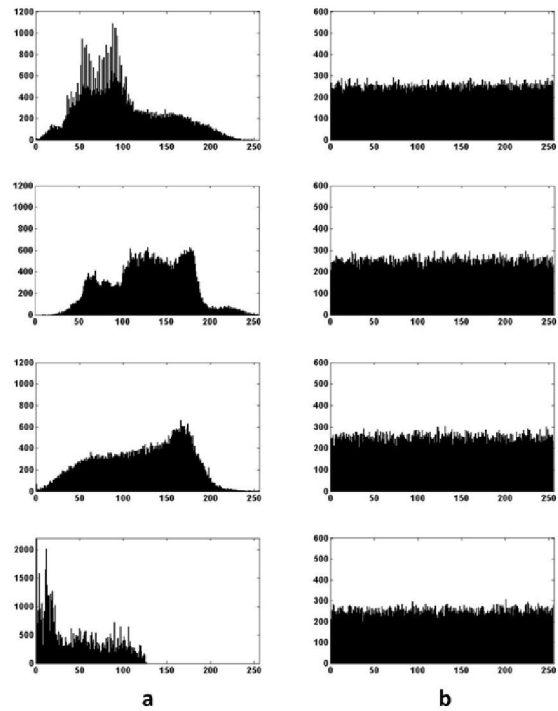

Fig.6. Histograms of images (Einstein, Ferdowsi, Rostam, and Pepper): (a) plain images; (b) encrypted images.

Table. 1. The average pixel intensity for plain images and encrypted images.

| Images | The average pixel intensity | |
|---|---|---|
| | Plain image | Encrypted image |
| Einstein | 98.05 | 127.15 |
| Ferdowsi | 130.15 | 127.81 |
| Rostam | 120.97 | 127.48 |
| Pepper | 43.28 | 127.98 |

### 4.3 Information Entropy

One of the characteristics of the randomness is entropy [35]. Information entropy is a mathematical entropy to measure the value of disorder of the storage and the data communication which defined by Claude E. Shannon in 1949 [29, 31]. The pixels of the encrypted image can have the number of different values (p). Suppose $N_i$ is the amount of pixels of the encrypted image that values are $0, 1, \ldots, p-1$, and also, the total amount of pixels of the encrypted image is N. The information entropy of the encrypted image is defined as

$$H = \sum_{i=0}^{p-1} \frac{N_i}{N} \log_2 \frac{N}{N_i}$$

in the interval $[0, \log_2 p]$. A method to scale the information entropy is normalized information entropy. In result, the normalized information entropy of the encrypted image is defined as

$$\bar{H} = \frac{\sum_{i=0}^{p-1} \frac{N_i}{N} \log_2 \frac{N}{N_i}}{\log_2 p}$$

in the interval [0,1]. The normalized entropy should be equal to one if the intensities of encrypted pixels had equal probability. In proposed method, the normalized entropy of encrypted images has values close to one, ranging from 0.9967 to 0.9999. Therefore, our encrypted images are nearly a random source with equal probability. Consequently, the proposed method is secure versus the entropy attack which agrees with the uniformity of histograms of the encrypted images (see Fig. 6).

## 4.4 Correlation Coefficient

In order to explain the values of uncorrelation between the pixels of encrypted image, the mathematical analysis performs on the encrypted image. The correlation coefficient analysis illustrates the correlation between two adjacent pixels in plain image and encrypted image. We randomly select 3000 pairs of two adjacent pixels (in horizontal, vertical, and diagonal direction) from plain image and encrypted image. The correlation coefficients are based on the following two equations, respectively (see Table 2) [30, 31]:

$$Cov(x,y) = \frac{1}{N} \sum_{i=1}^{N} (x_i - E(x))(y_i - E(y)),$$
$$r_{xy} = \frac{Cov(x,y)}{(D(x))^{\frac{1}{2}} (D(y))^{\frac{1}{2}}}$$

where

$$E(x) = \frac{1}{N} \sum_{i=1}^{N} (x_i), \quad D(x) = \frac{1}{N} \sum_{i=1}^{N} (x_i - E(x))^2$$

That the estimation of mathematical expectations of x is E(x), and the estimation of variance of x is D(x), and also the estimation of covariance between x and y is Cov(x,y). In addition, x and y in the image are grey scale values of two adjacent pixels [30, 33].

Table. 2. Correlation coefficients of two adjacent pixels in the plain images and the encrypted images.

| Direction | Plain Images | | | |
|---|---|---|---|---|
| | Einstein | Ferdowsi | Rostam | Pepper |
| Horizontal | 0.934 | 0.984 | 0.935 | 0.923 |
| Vertical | 0.963 | 0.946 | 0.947 | 0.911 |
| Diagonal | 0.940 | 0.937 | 0.969 | 0.918 |
| | Encrypted Images | | | |
| Horizontal | 0.0056 | 0.0027 | 0.0015 | 0.0043 |
| Vertical | 0.0031 | 0.0019 | 0.0045 | 0.0087 |
| Diagonal | 0.012 | 0.0056 | 0.0073 | 0.0042 |

## 4.5 Differential attack

Based on investigating value of sensitivity to the difference of two encrypted images, hackers always try to discover a relationship between the images [33]. Hackers apperceive the changes of the encrypted image based on the little change such as modifying one pixel of the plain image [32, 33]. They study the influence of one pixel change on the whole encrypted image by the proposed method, two current measures are used:
First, the number of pixels change rate (NPCR) that attempts for the number of pixels change rate while one pixel of plain image is changed. Second, the unified average changing intensity (UACI) that measures the average intensity of differences between the plain image and encrypted image.

The NPCR and The UACI are applied to test the influence of one pixel change on the whole of the image encrypted and can be defined as following [34]:

$$NPCR = \frac{\sum_{i,j} D(i,j)}{w \times h} \times 100\%$$

$$UACI = \frac{1}{w \times h} \left[ \sum_{i,j} \frac{C_1(i,j) - C_2(i,j)}{255} \right] \times 100\%$$

where h and w are the height and the width of $C_1$ or $C_2$. The $C_1$ and $C_2$ are two encrypted images whose corresponding plain images have the same size only and also have one pixel difference. The $C_1(i,j)$ and $C_2(i,j)$ are grey scale values of the pixels at grid $(i,j)$. The $D(i,j)$ determined by $C_1(i,j)$ and $C_2(i,j)$. If $C_1(i,j) = C_2(i,j)$, in result, $D(i,j) = 1$; otherwise, $D(i,j) = 0$. We have tried some tests on the proposed algorithm (256 grey scale image of size $256 \times 256$) to discover the scope of change made by one pixel change in the plain image (see Table 3). The results explain that the proposed algorithm can survive differential attack.

Table. 3. Results of the differential attack in the encrypted images (Minimum, maximum and average NPCR and UACI).

| Tests | Einstein | Ferdowsi | Rostam | Pepper |
|---|---|---|---|---|
| **NPCR%** | | | | |
| min: | 98.885 | 99.546 | 98.653 | 99.185 |
| max: | 99.125 | 98.876 | 98.654 | 98.249 |
| average: | 99.005 | 99.128 | 99.914 | 99.029 |
| **UACI%** | | | | |
| min: | 25.428 | 27.682 | 35.351 | 32.057 |
| max: | 25.519 | 27.271 | 35.954 | 31.817 |
| average: | 25.374 | 27.513 | 35.196 | 32.148 |

## 5. Conclusion

We have introduced a new method to image encryption. The proposed method used flexibility of the gradient haar wavelet transform and chaotic property of the rational order chaotic maps to produce the encrypted images. The method can be used in other fields such as image compression. The experiment results have illustrated the method which could be used in image encryption. Meantime, it is an efficient method for practical application in image processing and also real time transmission system.